\documentclass[%
reprint,
superscriptaddress,
amssymb,
aps,
prl,
floatfix,
]{revtex4-2}

\usepackage{graphicx}
\usepackage{rotating}
\usepackage{amsmath}
\usepackage{bbm}
\usepackage{subfigure}
\usepackage{color}
\usepackage{braket}
\usepackage{tikz}
\usepackage{hyperref}
\hypersetup{colorlinks, 
	linkcolor={blue!75!black!80!yellow},
	citecolor={blue!75!black!80!yellow}, 
	urlcolor={blue!75!black!80!yellow}
	}

\makeatletter \renewcommand\@make@capt@title[2]{%
\@ifx@empty\float@link{\@firstofone}{\expandafter\href\expandafter{\float@link}}%
\sffamily{\textbf{#1}}\@caption@fignum@sep#2 }

\begin{document}

\title{Simple Exchange-Correlation Energy Functionals for Strongly Coupled Light-Matter Systems based on the Fluctuation-Dissipation Theorem}

\author{Johannes Flick}
  \email[Electronic address:\;]{jflick@flatironinstitute.org}
\affiliation{Center for Computational Quantum Physics, Flatiron Institute, New York, NY, USA}

\date{\today}

\begin{abstract}
Recent experimental advances in strongly coupled light-matter systems has sparked the development of general \textit{ab-initio} methods capable of  describing interacting light-matter systems from first principles. One of these methods, quantum-electrodynamical density-functional theory (QEDFT), promises computationally efficient calculations for large  correlated light-matter systems with the quality of the calculation depending on the underlying approximation for the exchange-correlation functional. So far no true density-functional approximation has been introduced limiting the efficient application of the theory. In this paper, we introduce the first gradient-based density functional for the QEDFT exchange-correlation energy derived from the adiabatic-connection fluctuation-dissipation theorem. We benchmark this simple-to-implement approximation on small systems in optical cavities and  demonstrate its relatively low computational costs for fullerene molecules up to C$_{180}$ coupled to 400,000 photon modes in a dissipative optical cavity. This work now makes first principle calculations of much larger systems possible within the QEDFT framework effectively combining quantum optics with large-scale electronic structure theory. 
\end{abstract}

\date{\today}

\maketitle

\textit{Introduction}: The last few years have seen impressive experimental progress in achieving the strong coupling regime of light and matter. By strongly coupling molecular systems to optical cavities and nanoplasmonic structures, changes in several physical and chemical phenomena such as chemical reactivity for ground-state reactions~\cite{ebbesenTilting}, excited-state photochemical reactions~\cite{hutchison2012}, intermolecular vibrational energy transfer~\cite{Xiang2020}, harvesting of triplet excitons~\cite{Polak_2020}, two-photon phosphorescence~\cite{Ojambati2020}, ultrafast thermal modification~\cite{Liu2021}, among others,  have been demonstrated. At the same time, these experimental advances have motivated and inspired theoretical progress to introduce predictive methods that describe such systems from first principles. One of the first methods capable of describing strongly coupled light-matter systems from first principles was the generalization of density-functional theory (DFT)~\cite{kohn1965,kohn1999} dubbed quantum-electrodynamical density-functional theory (QEDFT)~\cite{tokatly2013,ruggenthaler2014}.  More recently also other methods from electronic structure theory and quantum chemistry have been extended to include strong light-matter interactions, such as Hartree-Fock wavefunction approaches~\cite{rivera2018, Haugland2020}, the multi-configuration time-dependent Hartree method~\cite{vendrell2018}, the exact factorization approaches~\cite{hoffmann2018, abedi2018}, and coupled-cluster theory~\cite{Mordovina2020,Haugland2020,Haugland2021,DePrince2021}, in particular the latter now allowing for highly accurate reference energies in this field.

In electronic structure theory and quantum chemistry, density-functional theory methods are very popular due to its low computational costs for large systems. The accuracy, quality, and numerical effort of these calculations strongly depends on the underlying approximation for the so-called exchange-correlation (xc) functional~\cite{kohn1999,burke2012}. Up to now, there has been a wide range of xc functionals proposed for electronic structure problems, many of which are now readily available in many DFT codes~\cite{Lehtola_2018}. Simple to evaluate approximations, such as the local density approximation (LDA)~\cite{hohenberg1964, perdew1981}, and its extension the generalized-gradient approximations (GGA)~\cite{Becke1988,perdew1996} have made DFT one of the most popular tool in electronic structure theory. Both of these classes of approximations are based on density functionals, meaning they are solely use the electronic density and spatial derivatives thereof as input. More expensive, but typically also more accurate orbitals functionals, such as hybrid functionals~\cite{heyd2003,Krukau2006} or OEP functionals~\cite{kuemmel2008}, include the Kohn-Sham orbitals explicitly, which prohibits their use for very large systems. 

At the same time the theory of QEDFT has not yet seen the same flexibility in terms of possible approximations for the electron-photon xc functional. Only one approximation has been developed so far, which is based on the OEP method~\cite{kuemmel2008} and is an orbital functional. Initially introduced in a formulation depending on both occupied and unoccupied Kohn-Sham orbitals~\cite{pellegrini2015}, this photon OEP functional has later been reformulated to a more computationally efficient scheme using Sternheimer equations~\cite{flick2018a} effectively circumventing any unoccupied orbitals. With computational costs comparable to electronic OEP functionals~\cite{kuemmel2008}, as an orbital functional, the method is rather expensive hindering application to large systems. Up to now, no simple to evaluate density functional for QEDFT has been introduced, which severely limits its applicability. In this paper we overcome this shortcoming of the QEDFT framework and introduce the first gradient-based density approximation for QEDFT. We do so in two steps: first we introduce the adiabatic-connection fluctuation-dissipation theorem of the QEDFT correlation energy, which allows us to express light-matter energy expressions in terms of response functions. In the second step, we replace the response function by dynamical polarizabilies and approximate these by density functionals~\cite{vydrov2010}. We exemplify our scheme on two smaller systems, the beryllium atom and the LiH molecule, inside optical cavities and compare the solution to reference energies. In addition, we simulate three different types of fullerenes, C$_{20}$,  C$_{60}$, and C$_{180}$ molecules coupled to a dissipative cavity described by an interaction with 400,000 photon modes.\\

\textit{Theory}: We start by discussing the Hamiltonian of general coupled light-matter systems in length-gauge and dipole approximation. Let us assume a system of $N_e$ interacting electrons  coupled to $N_p$ photon modes in an arbitrary electromagnetic environment. The Hamiltonian $\hat H$ of such a system is defined by~\cite{tokatly2013,ruggenthaler2014}
\begin{align}
\label{eqn:Hfull} 
\hat{H}&=\hat{T}_e + \hat{H}_p + \hat{H}_\text{int} + \hat{H}_\text{ext}\\
\hat{H}_\text{ext} &= \sum_{i=1}^{N_e}v_\text{ext}(\textbf{r}_i,t) + \sum_{\alpha=1}^{N_p} \frac{j^{(\alpha)}_\text{ext}(t)}{\omega_\alpha}\hat{q}_\alpha\\
\hat{H}_\text{int} &= -\omega_{\alpha}\hat{q}_{\alpha}{\boldsymbol{\lambda}_{\alpha}} \cdot \textbf{R}\nonumber\\
&+\frac{e^2}{4 \pi\epsilon_0}\sum_{i>j}^{N_e}\frac{1}{\left|\textbf{r}_i-\textbf{r}_j\right|} + \sum_{\alpha=1}^{N_p}\frac{1}{2} \left( \boldsymbol{\lambda}_{\alpha} \cdot \textbf R\right)^2,
\end{align}
where the electronic kinetic energy is given by $\hat T_e = -\sum_{i=1}^{N_e}  \frac{\hbar^2}{2m_e}\vec{\nabla}_i^2$ and $\hat{H}_p =\sum_{\alpha=1}^{N_p}\frac{1}{2}\left[\hat{p}^2_{\alpha}+\omega^2_{\alpha}\hat{q}^2_{\alpha}\right]$ describes the photonic Hamiltonian. The photonic operators $\hat q_{\alpha}$ and $\hat p_\alpha$ follow the usual quantum harmonic oscillator algebra and can be related to the physical quantities of the electric displacement operator and the magnetic field, respectively~\cite{pellegrini2015}. The external Hamiltonian $\hat{H}_\text{ext}$ includes the external electronic potential $v_\text{ext}(\textbf r,t)$ and external photonic current $j^{(\alpha)}_\text{ext}(t)$. All interactions in the system, i.e. electron-electron and electron-photon interactions are combined in $\hat{H}_\text{int}$. Electrons and photons couple via the electronic dipole moment $\textbf R = \sum_{i=1}^{N_e}\textbf{r}_i$ and the photon displacement operator $\hat{q}_\alpha$. As has been shown in earlier papers~\cite{tokatly2013,ruggenthaler2014,ruggenthaler2017groundstate}, the 1:1 correspondence of the set of internal (basic) variables, the electron density $n$ and the photon displacement coordinate $q_\alpha$, and the set of external variables, the electronic potential $v_\text{ext}$ and the external photonic current $j^{(\alpha)}_\text{ext}$, can be used to formulate a density-functional theory with the corresponding non-interacting Kohn-Sham system~\cite{tokatly2013,ruggenthaler2014,flick2015}. Based on these definitions, we can introduce the adiabatic connection between the physical interacting system and the non-interacting Kohn-Sham system with similar arguments as in e.g. Refs.~\cite{harris1984,hesselmann2011} as shown in appendix A.

Let us now turn our focus on the xc energy contributions. We can define the KS exchange energy with the Kohn-Sham ground-state wave function $\ket{\Psi_0}$ as follows
\begin{align}
E_x = \bra{\Psi_0}\hat{H}_\text{int}\ket{\Psi_0} - U ,  \label{eq:ex} 
\end{align}
where the classical energy $U$ can be defined by 
\begin{align*}
    U =&\iint d\textbf r  d\textbf r' \left[ \frac{e^2}{8\pi\epsilon_0}\frac{1}{|\textbf r -\textbf r'|}+\frac{1}{2} \sum_{\alpha=1}^{N_p}\left(\boldsymbol\lambda_\alpha\cdot\textbf r \right)\left(\boldsymbol\lambda_\alpha\cdot\textbf{r}'\right)\right]n(\textbf r)n(\textbf{r}')\\ +&\frac{1}{2}\sum_{\alpha=1}^{N_p}\int d\textbf r  \left(\boldsymbol\lambda_\alpha\cdot\textbf r \right)^2 n(\textbf r)
    \end{align*}
Now let us define the correlation energy $E_\text{c}(\gamma)$ using the adiabatic connection along the dimensionless parameter $\gamma \in [0,1]$. The parameter $\gamma$ interpolates between the noninteracting system ($\gamma=0$) and the physical Hamiltonian of Eq.~\ref{eqn:Hfull}  ($\gamma=1$). For more details we refer to appendix A. We find 
\begin{align*}
E_\text{c}(\gamma) = & \bra{\Psi_\gamma} \hat{T}_e + \hat{H}_p + \gamma\hat{H}_\text{int} \ket{\Psi_\gamma} \\
&- \bra{\Psi_0} \hat{T}_e + \hat{H}_p + \gamma\hat{H}_\text{int} \ket{\Psi_0}.\nonumber
\end{align*}
Using $\frac{dE_c(\gamma)}{d\gamma}$ allows us to formulate the physical correlation energy as
\begin{align*}
E_c =& \int_{0}^1 d\gamma \frac{dE_c(\gamma)}{d\gamma}.
\end{align*}
As a next step, let us formulate the adiabatic-connection fluctuation-dissipation theorem of the correlation energy $E_c$, which we define in terms of electron-photon response functions. These response functions have been previously introduced in Ref.~\cite{flick2019} and are defined e.g. as

\begin{align*}
\chi^{q_\alpha}_n(\textbf{r},\nu) =& \sum_{k\neq 0} \frac{\bra{\Phi_0}\hat{n}(\textbf{r})\ket{\Phi_k}\bra{\Phi_k}\hat{q}_\alpha\ket{\Phi_0}}{E_0-E_k + \nu} \\
&+ \frac{\bra{\Phi_k}\hat{n}(\textbf{r})\ket{\Phi_0}\bra{\Phi_0}\hat{q}_\alpha\ket{\Phi_k}}{E_0-E_k - \nu},\nonumber
\end{align*}
with complex frequencies $\nu=\omega+i\eta$. $\eta$ describes here a convergence factor and $\eta\rightarrow 0$ for all practical purposes.  We note that the real continuous frequencies $\omega$ used in $\nu$ are different to the discrete frequencies of the photon modes $\omega_\alpha$ used in Eq.~\ref{eqn:Hfull} distinguished by the additional subindex $\alpha$. Further, $\ket{\Phi_0}$ is here now the ground-state wave-function with energy $E_0$ to the Hamiltonian in Eq.~\ref{eqn:Hfull} and $\ket{\Phi_k}$ describes an excited state of energy $E_k$. It is straightforward to generalize $\chi^{q_\alpha}_n(\textbf{r},\nu)$ to $\chi^{q_\alpha}_{n,\gamma}(\textbf{r},\nu)$ by using the corresponding set of eigenstates of the adiabatic-connection Hamiltonian $\hat{H}_\gamma$ discussed in appendix A.

We can now connect the correlation energy and these response functions via the adiabatic connection relation as defined above and find the following equations
\begin{align}
\label{eq:fluct-diss1}
E_\text{c}^{(1)} = \frac{1}{2\pi} \int^1_0 d\gamma  &\int d\textbf{r} \sum_{\alpha=1}^{N_p} \omega_\alpha\left({\boldsymbol \lambda_\alpha}\cdot \textbf{r}\right) \\
&\times\int_0^\infty d\omega\left[\chi^{q_\alpha}_{n,\gamma}(\textbf{r},i\omega) - \chi^{q_\alpha}_{n,0}(\textbf{r},i\omega)\right] \nonumber
\end{align}
\begin{align}
\label{eq:fluct-diss2}
E_\text{c}^{(2)} &= \\
-& \frac{1}{2\pi} \int^1_0 d\gamma  \int d\textbf{r} \left[ \frac{e^2}{4\pi\epsilon_0}\frac{1}{|\textbf r -\textbf r'|}+\sum_{\alpha=1}^{N_p} \left(\boldsymbol\lambda_\alpha\cdot\textbf r \right)\left(\boldsymbol\lambda_\alpha\cdot\textbf{r}'\right) \right] \nonumber\\
&\times\int_0^\infty d\omega\left[\chi^{n}_{n,\gamma}(\textbf{r},i\omega) - \chi^{n}_{n,0}(\textbf{r},i\omega)\right] \nonumber
\end{align}
where $\chi^{n}_{n}$ are the regular density-density response functions~\cite{hesselmann2011,flick2019} and, with $E_c = E_c^{(1)} + E_c^{(2)}$, leads to the adiabatic-connection fluctuation-dissipation theorem of the correlation energy. Here, $E_c^{(1)}$ includes all correlation effects of the explicit electron-photon interaction stemming from the $\left( \boldsymbol\lambda_\alpha \hat q_\alpha \cdot \textbf R\right)$ part in Eq.~\ref{eqn:Hfull}, while $E_c^{(2)}$ includes all correlation effects due to the electron-electron interaction and the $\left(\boldsymbol\lambda_\alpha \cdot \textbf R\right)^2$ term, also called the $R^2$-term.

In the remaining part of this paper, we now will use these exact expressions of Eqs.~\ref{eq:fluct-diss1} and \ref{eq:fluct-diss2} as starting points for approximations to $E_c$. Having now a connection between the correlation energies in terms of light-matter response functions at hand, we can use the Dyson equations defined in Ref.~\cite{flick2019} to connect the interacting response functions to Kohn-Sham response functions. One simple approximation for Eq.~\ref{eq:fluct-diss1} can be obtained by replacing the integral over $\gamma$ by a sum including only the boundary points $\gamma=1$, and $\gamma=0$, where we then use 
\begin{align*}
    \chi^{q_\alpha}_{n,\gamma =1}(\textbf{r},i\omega) =\sum_\beta \chi^{q_\alpha}_{q_\beta,s}g^{n_\beta}_M \chi_{n}^{n}\longrightarrow \sum_\beta \chi^{q_\alpha}_{q_\beta,s}g^{n_\beta}_M \chi_{n_s}^{n},
\end{align*}
with the photon-photon response function $\chi^{q_\alpha}_{q_\beta,s}$ and the kernel $g^{n_\beta}_M$ defined as in Ref.~\cite{flick2019}. The additional label 's' refers to the corresponding quantity in the Kohn-Sham system. By further introducing the dynamic polarizability $\alpha_{\mu\nu}$ to these equations, we find the following expressions

\begin{align}
E_\text{c}^{(1,1)} &=\frac{1}{2\pi}\sum_{\alpha=1}^{N_p}\int_0^\infty d\omega \; \frac{\omega_\alpha^2}{\omega^2+\omega_\alpha^2} \boldsymbol  \lambda_\alpha \cdot \boldsymbol \alpha(i\omega) \cdot \boldsymbol\lambda_\alpha\label{eq:exc2}
\end{align}

with
\begin{align}
\alpha_{\mu\nu}(i\omega) &= -2\sum_{ia} \frac{\left(\epsilon_a-\epsilon_i\right)\bra{\varphi_a}r_\mu\ket{\varphi_i}\bra{\varphi_i}r_\nu\ket{\varphi_a}}{\left(\epsilon_a-\epsilon_i\right)^2 + \omega^2},
\end{align}
where the indices $i$, and $a$ refer to occupied and unoccupied orbitals, respectively. If we combine this result with the exchange energy of Eq.~\ref{eq:ex}, without considering the electron-electron interaction, 
\begin{align}
\label{eq:exc1}
    E_x &= -\frac{1}{2\pi}\sum_{\alpha=1}^{N_p}\int_0^\infty d\omega\; \boldsymbol  \lambda_\alpha \cdot \boldsymbol \alpha(i\omega) \cdot \boldsymbol\lambda_\alpha
\end{align}
the resulting energy expression $E_x^{(p)} = E_x + E_c^{(1,1)}$ becomes identical to the energy expression considered in the photon OEP approach introduced in Refs.~\cite{pellegrini2015,flick2018a} and we will refer to it in the following as electron-photon exchange energy. We show this connection explicitly in the appendix B. $E_x$ and $E_c^{(1,1)}$ are here combined with the argument of both being $\sim \lambda_\alpha^2$. The lowest order contributions from $E_c^{(2)}$ in Eq.~\ref{eq:fluct-diss2} are already proportional to $\sim \lambda_\alpha^4$ and will be therefore neglected. We note however that including higher order effects using Eqs.~\ref{eq:fluct-diss1}-\ref{eq:fluct-diss2} is straightforward. Physically such an approximation corresponds to including one-photon exchange processes explicitly, while neglecting higher order processes~\cite{pellegrini2015,flick2018a}.\\

In a last step to transform Eq.~\ref{eq:exc2} and ~\ref{eq:exc1} into a density-functional, we approximate the dynamic polarizability $\alpha_{\mu\nu}$ by the VV10 polarizability introduced in Ref.~\cite{vydrov2010}. One of the main assumptions is isotropy of the physical system. This approximation reads:
\begin{align*}
    \alpha(iu) = \frac{1}{4\pi}\int d\textbf{r}\frac{\omega_p^2(\textbf r)}{\omega_p^2(\textbf r)/3 + \omega_g^2(\textbf r) + u^2}
\end{align*}
with the plasmon frequency of the free electron gas $\omega_p^2(\textbf r) = 4\pi n(\textbf{r})e^2/m$ and the gap frequency $\omega_g^2(\textbf r) = C \frac{\hbar^2}{m^2}\left|\frac{\nabla n(\textbf r)}{n(\textbf r)} \right|^4$, where $C$ is typically chosen as $C = 0.0089$~\cite{vydrov2010}. The plasmon frequency $\omega_p$ has a functional dependency on the electron density $n$, and the gap frequency $\omega_g$ on $n$ and its gradient $\nabla n$. We also note that other approximations for the dynamic polarizability without including the gradient of the density have been derived, but have been shown to be substantially less accurate~\cite{vydrov2010, Hermann2017}.

Combining all those equations, we can reformulate Eqs.~\ref{eq:exc2} and \ref{eq:exc1} as
\begin{align}
E_\text{x}^\text{(GA)}[n,\nabla n] & = \frac{1}{16\pi}\sum_{\alpha=1}^{N_p}|\boldsymbol\lambda_\alpha|^2\int d\textbf{r}\frac{\omega_p^2(\textbf{r})}{ \sqrt{\omega_p^2(\textbf r)/3+\omega_g^2(\textbf r)}+\omega_\alpha}.
\label{eq:gga}
\end{align}

This equation now yields an approximation of the electron-photon exchange energy depending on the electron density and its spatial derivatives through the plasma frequency $\omega_p$ and the gap frequency $\omega_g$ and can be easily paired with any other functional that describes the electronic structure~\cite{Lehtola_2018}.

\textit{Application}: In the remaining part of the paper, we exemplify the proposed scheme on two different setups. We will use the energy expression of Eq.~\ref{eq:gga} in a non-self-consistent form. In the first setup, we compare the approximate energy expression of Eq.~\ref{eq:gga} to the energy expression of the self-consistent photon OEP energy of Refs.~\cite{pellegrini2015,flick2018a} to which it is an approximation. We also compare to the KLI approximation~\cite{flick2018a}, which solves the OEP equations in an approximate form and include a comparison to a non-self-consistent solution of the OEP equation (OEP-ss). These calculations are performed with strong coupling to a single photon mode for a single beryllium atom and the LiH molecule. These two examples serve as a good test case, since the beryllium atom is a perfectly symmetric electronic system and the LiH molecule being a dimer is anisotropic. In the second setup, we study a much larger system, where the solution of the full photon OEP formalism is not possible. We compare the effects of cavity dissipation on the ground-state energy for three fullerenes, C$_{20}$, C$_{60}$, and C$_{180}$. The computational details of these calculations are described in the appendix C.

\begin{figure}[t]
\centering
\includegraphics[width=1.0\linewidth]{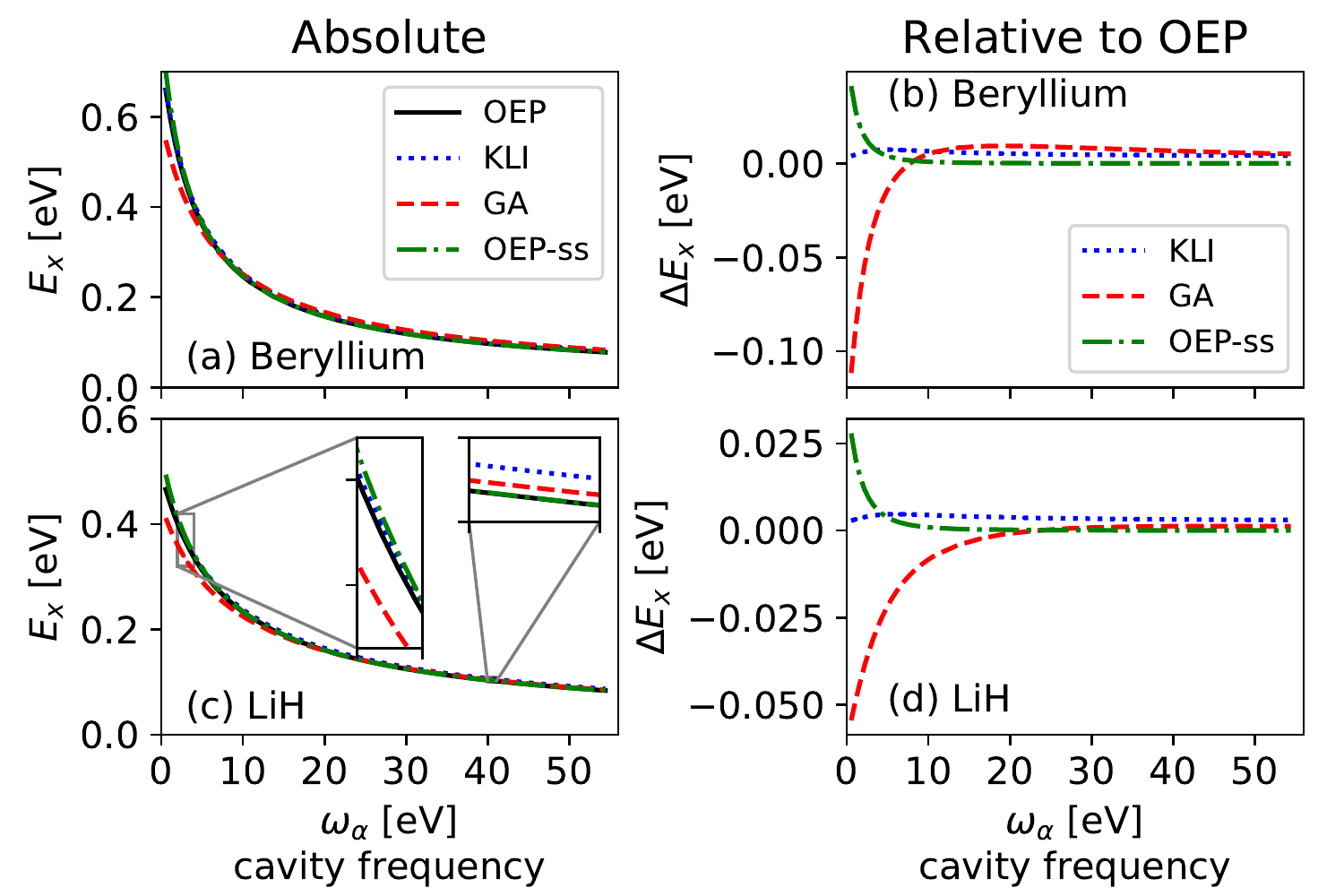}
\caption{\textbf{Comparison of the different schemes for the beryllium atom and the LiH molecule.} In (a) we show the absolute energy scale of the KLI (black), OEP (blue), GA (red), and OEP-ss (single shot) (green). (b) show the difference of KLI (black), GA (red), and OEP-ss (single shot) (green) with respect to the ptOEP energies. Same for the LiH molecule in (c) and (d), respectively.}
\label{fig:01}
\end{figure}

We first discuss the benchmark analysis shown in Fig.~\ref{fig:01}, where we test the accuracy of the developed approximate energy functional. We show the results for the system of a single beryllium atom and the LiH molecule, both coupled strongly with $\lambda_\alpha=0.1$ a.u. to an optical cavity mode of variable frequency. In Fig.~\ref{fig:01} (a) and (c), we show the absolute values of the electron-photon $E_{xc}$ for different levels of theory, i.e. KLI (black), OEP (blue), GA (red), and OEP-ss (green),
for beryllium and LiH, respectively. The KLI and OEP schemes are discussed in Ref.~\cite{flick2018a} and are self-consistent solutions with this level of theory applied to the electronic and photon degrees of freedom. For the OEP-ss (single-shot) scheme, we calculate the electronic OEP solution and add the photon OEP energy in a single shot, without additional self-consistency. All curves show very similar behavior highlighting the accuracy of the individual approximations. In more detail this can be also seen in (b) and (d), where we plot the difference of the KLI (black), GA (red), and OEP-ss (green) scheme to the OEP energy. We find the most accurate energy at this level of theory is the KLI approximation compared to the OEP energy, which can be expected, since the KLI approximation is using the same expression for the xc energy as the OEP functional~\cite{flick2018a}. The OEP-ss shows for the low frequency regime smaller deviations ~5\%, which become more pronounced in the GA scheme. We note that this larger deviation of the GA scheme is in agreement with the quality of the dynamic polarizability of the VV10 approximation demonstrated in Ref.~\cite{vydrov2010}. These findings can be expected, since the energy $E^\text{(GA)}_\text{x}$ discussed in Eq.~\ref{eq:gga} is an approximation to the OEP-ss energy. Overall we find that the GA approximation gives accurate energies for the two studied systems with deviations smaller than 10\% for both the isotropic and anisotropic system, but dramatically less computational costs than OEP, KLI, and OEP-ss.

\begin{figure}[t]
\centering
\includegraphics[width=1.0\linewidth]{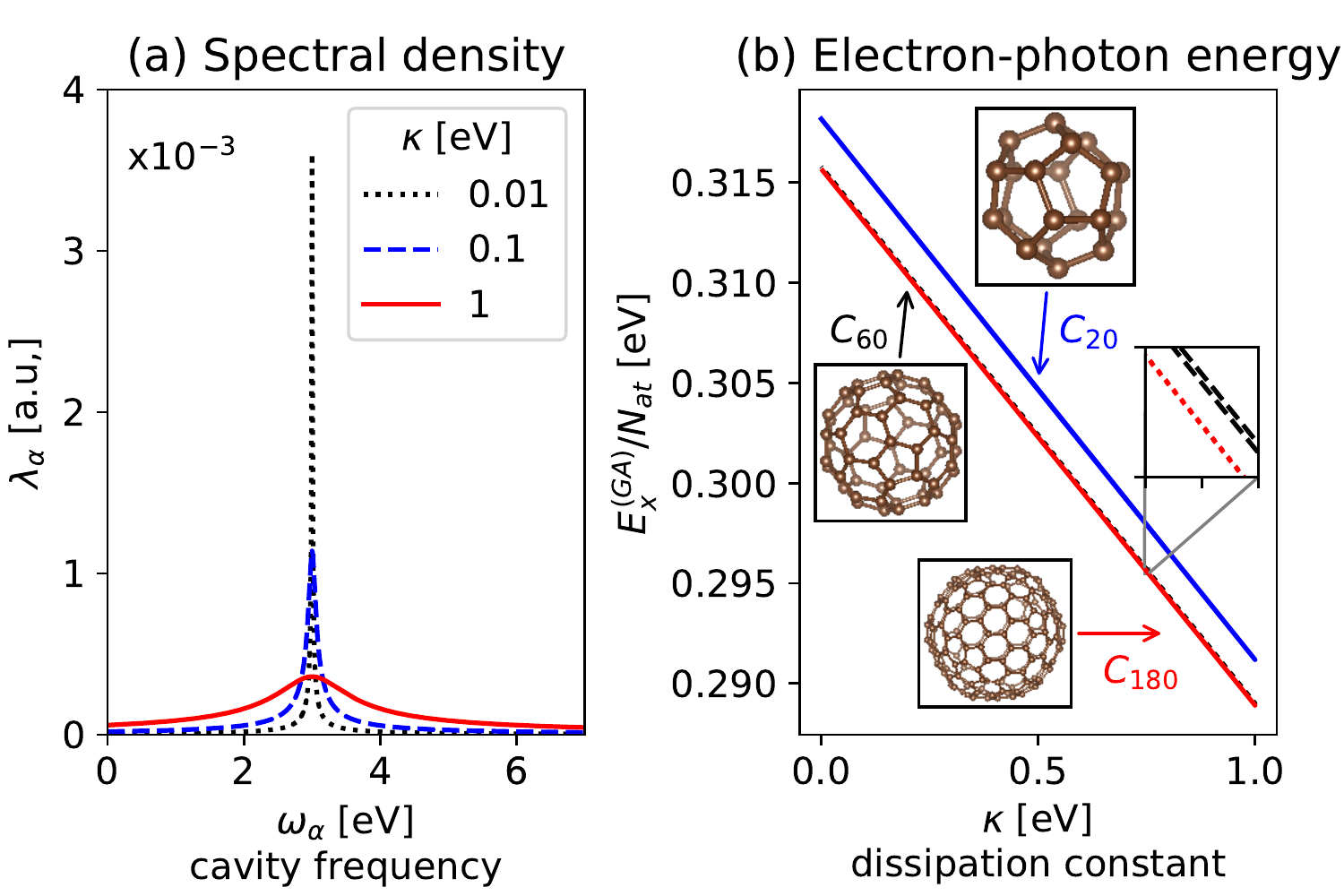}
\caption{\textbf{Spectral densities and electron-photon energy for C$_{20}$, $C_{60}$, and $C_{180}$.} In (a), we show the spectral density with different $\lambda_k$ values over cavity frequency $\omega_\alpha$ for $\kappa=0.01$ eV in black, $\kappa=0.1$ eV in blue, and $\kappa=1$ eV in red. (b) shows the electron-photon energy $E^\text{(GA)}_\text{xc}/N_{at}$ for C$_{20}$ (blue),  C$_{60}$ (black), and $C_{180}$ (red) for different dissipation constants $\kappa \in [0,1]$ eV.}
\label{fig:02}
\end{figure}

In the second example shown in Fig.~\ref{fig:02}, we show three fullerene molecules, C$_{20}$, C$_{60}$, and C$_{180}$, in an lossy optical cavity and study how the losses affect the electron-photon interaction in the ground state. To describe the losses, we follow the scheme introduced in Ref.~\cite{wang2021}. We assume one strongly coupled photon mode, but in contrast to the previous example, where the cavity produced a sharp single photon mode as resonance, in this case we broaden it by explicitly including 400,000 photon modes. The values for $\lambda_\alpha$ are sampled by ~\cite{hummer2013,wang2021} $|\boldsymbol{\lambda}_\alpha|^2=|\boldsymbol{\lambda}_{\rm c}|^2L(\Delta\omega,\kappa,\omega_{c})$,
with the Lorentizan profile function $L(\Delta\omega,\kappa,\omega_{c})=\Delta\omega\frac{1}{2\pi}\frac{\kappa}{(\omega_{\alpha}-\omega_c)^2+(\kappa/2)^2}$ defined by the broadening and dissipation constant $\kappa$.

In our example, we choose $\omega_c = 3$ eV and $\lambda_c = 0.1$ a.u. The resulting spectral densities for a subset of the calculated dissipation constants are shown in Fig.~\ref{fig:02} (a) with the distributions centered around $\omega_c = 3$ eV. We show three different values for with $\kappa =$ 0.01 eV (black), 0.1 eV (blue), and 1 eV (red) and find that the distribution becomes broader and has a lower the maximal value for higher $\kappa$ values due to the Lorentzian function. Using these spectral densities as input, we plot the electron-photon exchange energy $E^{(p)}_{x}$ normalized to the number of atoms $N_\text{at}$ for the three fullerene systems in Fig.~\ref{fig:02} (b). We also plot the value of $\kappa=0$ eV, which is equal to including only a single photon mode with $\omega_c$, and $\lambda_c$ in the calculation. We find a linear behavior of $E^{(GA)}_{x}/N_\text{at}$ with $\kappa$. This linear behavior originates from the connection of $\kappa$ to the lifetimes $\Gamma$ in the system, which is then connected to the Lamb shift via the Kramers-Kronig relation~\cite{scheel2009macroscopic}. Interestingly we find different offset values in the linear relation for the three different systems, also see inset. We find the offset to be larger for smaller systems, with C$_{20}$ (blue curve) having the largest offset and a similar offset for C$_{60}$ (black curve) and C$_{180}$ (red curve), but smallest for $C_{180}$ (inset). This illustrates the dependency of the electron-photon exchange energy on the electronic structure of the system. We also note quite sizable values of around 300 meV per atom highlighting the importance of this contribution for larger systems.

\textit{Summary and Conclusion}: The presented work introduces a general reliable and fast approximate scheme for efficient calculations of strongly coupled light-matter systems. By formulating the adiabatic-connection fluctuation-dissipation theorem for the correlation energy, we are able to introduce a simple approximation scheme for this quantity. Formulating response functions in terms of dynamic polarizabilities and approximating those using the VV10 scheme allows us to construct an energy functional only using the electron density and spatial derivatives. The developed functional is straightforward to implement and relatively inexpensive. Our calculations are the first \textit{ab initio} calculations that use a gradient-based density functional for the xc energy and we demonstrate the accuracy on the beryllium atom and the LiH molecule strongly coupled to a photon mode. In an additional example, we simulate  fullerene molecules up to C$_{180}$ coupled to 400,000 photonic modes in an dissipative optical cavity, which makes it the largest QEDFT calculation with an xc functional so far. There are several extensions possible to this new route of constructing xc functionals. While in this work, we used a simple termination of Eq.~\ref{eq:fluct-diss1} at the first order, which corresponds to including single-photon exchange processes~\cite{pellegrini2015, flick2018a}, extending this scheme to higher order to include two-, three- and higher order processes is straightforward. One possible way to construct the full response function would be to use the photon Casida equation~\cite{flick2019}, e.g. for accurate benchmarks. We expect in particular two-photon processes to become important for systems that show strong dispersion interactions~\cite{keller2012quantum}. The study of effects of anisotropy, which is currently not included, or other approximate schemes for the dynamical polarizability~\cite{Hermann2017} will be an important next step to increase the accuracy of the calculations. Additionally, the introduced energy functional can also be used to derive the corresponding $v_\text{xc}$, which will allow for self-consistent solutions. We also want to highlight the connection of our work to previous work on van-der-Waals systems and Casimir interactions~\cite{venkataram2017}. We envision that the introduced new route for efficient approximation will establish QEDFT as the toolbox to understand the ground-state behavior of strongly coupled light-matter systems and study altering properties of molecular and extended systems by accessing light-matter correlations.

\textit{Acknowledgements}: I would like to thank Derek Wang, Michael Ruggenthaler, Angel Rubio, Nicolas Rivera, and Prineha Narang for insightful discussions. All calculations were performed using the computational facilities of the Flatiron Institute. The Flatiron Institute is a division of the Simons Foundation.

\bibliography{refs} 

\begin{thebibliography}{46}%
\makeatletter
\providecommand \@ifxundefined [1]{%
 \@ifx{#1\undefined}
}%
\providecommand \@ifnum [1]{%
 \ifnum #1\expandafter \@firstoftwo
 \else \expandafter \@secondoftwo
 \fi
}%
\providecommand \@ifx [1]{%
 \ifx #1\expandafter \@firstoftwo
 \else \expandafter \@secondoftwo
 \fi
}%
\providecommand \natexlab [1]{#1}%
\providecommand \enquote  [1]{``#1''}%
\providecommand \bibnamefont  [1]{#1}%
\providecommand \bibfnamefont [1]{#1}%
\providecommand \citenamefont [1]{#1}%
\providecommand \href@noop [0]{\@secondoftwo}%
\providecommand \href [0]{\begingroup \@sanitize@url \@href}%
\providecommand \@href[1]{\@@startlink{#1}\@@href}%
\providecommand \@@href[1]{\endgroup#1\@@endlink}%
\providecommand \@sanitize@url [0]{\catcode `\\12\catcode `\$12\catcode
  `\&12\catcode `\#12\catcode `\^12\catcode `\_12\catcode `\%12\relax}%
\providecommand \@@startlink[1]{}%
\providecommand \@@endlink[0]{}%
\providecommand \url  [0]{\begingroup\@sanitize@url \@url }%
\providecommand \@url [1]{\endgroup\@href {#1}{\urlprefix }}%
\providecommand \urlprefix  [0]{URL }%
\providecommand \Eprint [0]{\href }%
\providecommand \doibase [0]{https://doi.org/}%
\providecommand \selectlanguage [0]{\@gobble}%
\providecommand \bibinfo  [0]{\@secondoftwo}%
\providecommand \bibfield  [0]{\@secondoftwo}%
\providecommand \translation [1]{[#1]}%
\providecommand \BibitemOpen [0]{}%
\providecommand \bibitemStop [0]{}%
\providecommand \bibitemNoStop [0]{.\EOS\space}%
\providecommand \EOS [0]{\spacefactor3000\relax}%
\providecommand \BibitemShut  [1]{\csname bibitem#1\endcsname}%
\let\auto@bib@innerbib\@empty
\bibitem [{\citenamefont {Thomas}\ \emph {et~al.}(2019)\citenamefont {Thomas},
  \citenamefont {Lethuillier-Karl}, \citenamefont {Nagarajan}, \citenamefont
  {Vergauwe}, \citenamefont {J.~George}, \citenamefont {Shalabney},
  \citenamefont {Devaux}, \citenamefont {Genet}, \citenamefont {Moran},\ and\
  \citenamefont {Ebbesen}}]{ebbesenTilting}%
  \BibitemOpen
  \bibfield  {author} {\bibinfo {author} {\bibfnamefont {A.}~\bibnamefont
  {Thomas}}, \bibinfo {author} {\bibfnamefont {L.}~\bibnamefont
  {Lethuillier-Karl}}, \bibinfo {author} {\bibfnamefont {K.}~\bibnamefont
  {Nagarajan}}, \bibinfo {author} {\bibfnamefont {R.~M.~A.}\ \bibnamefont
  {Vergauwe}}, \bibinfo {author} {\bibfnamefont {T.~C.}\ \bibnamefont
  {J.~George}}, \bibinfo {author} {\bibfnamefont {A.}~\bibnamefont
  {Shalabney}}, \bibinfo {author} {\bibfnamefont {E.}~\bibnamefont {Devaux}},
  \bibinfo {author} {\bibfnamefont {C.}~\bibnamefont {Genet}}, \bibinfo
  {author} {\bibfnamefont {J.}~\bibnamefont {Moran}},\ and\ \bibinfo {author}
  {\bibfnamefont {T.~W.}\ \bibnamefont {Ebbesen}},\ }\bibfield  {title}
  {\bibinfo {title} {Tilting a ground-state reactivity landscape by vibrational
  strong coupling},\ }\href
  {https://science.sciencemag.org/content/363/6427/615.abstract} {\bibfield
  {journal} {\bibinfo  {journal} {Science}\ }\textbf {\bibinfo {volume}
  {364}},\ \bibinfo {pages} {615} (\bibinfo {year} {2019})}\BibitemShut
  {NoStop}%
\bibitem [{\citenamefont {Hutchison}\ \emph {et~al.}(2012)\citenamefont
  {Hutchison}, \citenamefont {Schwartz}, \citenamefont {Genet}, \citenamefont
  {Devaux},\ and\ \citenamefont {Ebbesen}}]{hutchison2012}%
  \BibitemOpen
  \bibfield  {author} {\bibinfo {author} {\bibfnamefont {J.~A.}\ \bibnamefont
  {Hutchison}}, \bibinfo {author} {\bibfnamefont {T.}~\bibnamefont {Schwartz}},
  \bibinfo {author} {\bibfnamefont {C.}~\bibnamefont {Genet}}, \bibinfo
  {author} {\bibfnamefont {E.}~\bibnamefont {Devaux}},\ and\ \bibinfo {author}
  {\bibfnamefont {T.~W.}\ \bibnamefont {Ebbesen}},\ }\bibfield  {title}
  {\bibinfo {title} {Modifying chemical landscapes by coupling to vacuum
  fields},\ }\href {https://doi.org/10.1002/anie.201107033} {\bibfield
  {journal} {\bibinfo  {journal} {Angew. Chem. Int. Ed.}\ }\textbf {\bibinfo
  {volume} {51}},\ \bibinfo {pages} {1592} (\bibinfo {year}
  {2012})}\BibitemShut {NoStop}%
\bibitem [{\citenamefont {Xiang}\ \emph {et~al.}(2020)\citenamefont {Xiang},
  \citenamefont {Ribeiro}, \citenamefont {Du}, \citenamefont {Chen},
  \citenamefont {Yang}, \citenamefont {Wang}, \citenamefont {Yuen-Zhou},\ and\
  \citenamefont {Xiong}}]{Xiang2020}%
  \BibitemOpen
  \bibfield  {author} {\bibinfo {author} {\bibfnamefont {B.}~\bibnamefont
  {Xiang}}, \bibinfo {author} {\bibfnamefont {R.~F.}\ \bibnamefont {Ribeiro}},
  \bibinfo {author} {\bibfnamefont {M.}~\bibnamefont {Du}}, \bibinfo {author}
  {\bibfnamefont {L.}~\bibnamefont {Chen}}, \bibinfo {author} {\bibfnamefont
  {Z.}~\bibnamefont {Yang}}, \bibinfo {author} {\bibfnamefont {J.}~\bibnamefont
  {Wang}}, \bibinfo {author} {\bibfnamefont {J.}~\bibnamefont {Yuen-Zhou}},\
  and\ \bibinfo {author} {\bibfnamefont {W.}~\bibnamefont {Xiong}},\ }\bibfield
   {title} {\bibinfo {title} {Intermolecular vibrational energy transfer
  enabled by microcavity strong light{\textendash}matter coupling},\ }\href
  {https://doi.org/10.1126/science.aba3544} {\bibfield  {journal} {\bibinfo
  {journal} {Science}\ }\textbf {\bibinfo {volume} {368}},\ \bibinfo {pages}
  {665} (\bibinfo {year} {2020})}\BibitemShut {NoStop}%
\bibitem [{\citenamefont {Polak}\ \emph {et~al.}(2020)\citenamefont {Polak},
  \citenamefont {Jayaprakash}, \citenamefont {Lyons}, \citenamefont
  {Mart{\'{\i}}nez-Mart{\'{\i}}nez}, \citenamefont {Leventis}, \citenamefont
  {Fallon}, \citenamefont {Coulthard}, \citenamefont {Bossanyi}, \citenamefont
  {Georgiou}, \citenamefont {Anthony J.~Petty}, \citenamefont {Anthony},
  \citenamefont {Bronstein}, \citenamefont {Yuen-Zhou}, \citenamefont
  {Tartakovskii}, \citenamefont {Clark},\ and\ \citenamefont
  {Musser}}]{Polak_2020}%
  \BibitemOpen
  \bibfield  {author} {\bibinfo {author} {\bibfnamefont {D.}~\bibnamefont
  {Polak}}, \bibinfo {author} {\bibfnamefont {R.}~\bibnamefont {Jayaprakash}},
  \bibinfo {author} {\bibfnamefont {T.~P.}\ \bibnamefont {Lyons}}, \bibinfo
  {author} {\bibfnamefont {L.~{\'{A}}.}\ \bibnamefont
  {Mart{\'{\i}}nez-Mart{\'{\i}}nez}}, \bibinfo {author} {\bibfnamefont
  {A.}~\bibnamefont {Leventis}}, \bibinfo {author} {\bibfnamefont {K.~J.}\
  \bibnamefont {Fallon}}, \bibinfo {author} {\bibfnamefont {H.}~\bibnamefont
  {Coulthard}}, \bibinfo {author} {\bibfnamefont {D.~G.}\ \bibnamefont
  {Bossanyi}}, \bibinfo {author} {\bibfnamefont {K.}~\bibnamefont {Georgiou}},
  \bibinfo {author} {\bibfnamefont {I.}~\bibnamefont {Anthony J.~Petty}},
  \bibinfo {author} {\bibfnamefont {J.}~\bibnamefont {Anthony}}, \bibinfo
  {author} {\bibfnamefont {H.}~\bibnamefont {Bronstein}}, \bibinfo {author}
  {\bibfnamefont {J.}~\bibnamefont {Yuen-Zhou}}, \bibinfo {author}
  {\bibfnamefont {A.~I.}\ \bibnamefont {Tartakovskii}}, \bibinfo {author}
  {\bibfnamefont {J.}~\bibnamefont {Clark}},\ and\ \bibinfo {author}
  {\bibfnamefont {A.~J.}\ \bibnamefont {Musser}},\ }\bibfield  {title}
  {\bibinfo {title} {Manipulating molecules with strong coupling: harvesting
  triplet excitons in organic exciton microcavities},\ }\href
  {https://doi.org/10.1039/c9sc04950a} {\bibfield  {journal} {\bibinfo
  {journal} {Chemical Science}\ }\textbf {\bibinfo {volume} {11}},\ \bibinfo
  {pages} {343} (\bibinfo {year} {2020})}\BibitemShut {NoStop}%
\bibitem [{\citenamefont {Ojambati}\ \emph {et~al.}(2020)\citenamefont
  {Ojambati}, \citenamefont {Chikkaraddy}, \citenamefont {Deacon},
  \citenamefont {Huang}, \citenamefont {Wright},\ and\ \citenamefont
  {Baumberg}}]{Ojambati2020}%
  \BibitemOpen
  \bibfield  {author} {\bibinfo {author} {\bibfnamefont {O.~S.}\ \bibnamefont
  {Ojambati}}, \bibinfo {author} {\bibfnamefont {R.}~\bibnamefont
  {Chikkaraddy}}, \bibinfo {author} {\bibfnamefont {W.~M.}\ \bibnamefont
  {Deacon}}, \bibinfo {author} {\bibfnamefont {J.}~\bibnamefont {Huang}},
  \bibinfo {author} {\bibfnamefont {D.}~\bibnamefont {Wright}},\ and\ \bibinfo
  {author} {\bibfnamefont {J.~J.}\ \bibnamefont {Baumberg}},\ }\bibfield
  {title} {\bibinfo {title} {Efficient generation of two-photon excited
  phosphorescence from molecules in plasmonic nanocavities},\ }\href
  {https://doi.org/10.1021/acs.nanolett.0c01593} {\bibfield  {journal}
  {\bibinfo  {journal} {Nano Letters}\ }\textbf {\bibinfo {volume} {20}},\
  \bibinfo {pages} {4653} (\bibinfo {year} {2020})}\BibitemShut {NoStop}%
\bibitem [{\citenamefont {Liu}\ \emph {et~al.}(2021)\citenamefont {Liu},
  \citenamefont {Menon},\ and\ \citenamefont {Sfeir}}]{Liu2021}%
  \BibitemOpen
  \bibfield  {author} {\bibinfo {author} {\bibfnamefont {B.}~\bibnamefont
  {Liu}}, \bibinfo {author} {\bibfnamefont {V.~M.}\ \bibnamefont {Menon}},\
  and\ \bibinfo {author} {\bibfnamefont {M.~Y.}\ \bibnamefont {Sfeir}},\
  }\bibfield  {title} {\bibinfo {title} {Ultrafast thermal modification of
  strong coupling in an organic microcavity},\ }\href
  {https://doi.org/10.1063/5.0031560} {\bibfield  {journal} {\bibinfo
  {journal} {APL Photonics}\ }\textbf {\bibinfo {volume} {6}},\ \bibinfo
  {pages} {016103} (\bibinfo {year} {2021})}\BibitemShut {NoStop}%
\bibitem [{\citenamefont {Kohn}\ and\ \citenamefont {Sham}(1965)}]{kohn1965}%
  \BibitemOpen
  \bibfield  {author} {\bibinfo {author} {\bibfnamefont {W.}~\bibnamefont
  {Kohn}}\ and\ \bibinfo {author} {\bibfnamefont {L.~J.}\ \bibnamefont
  {Sham}},\ }\bibfield  {title} {\bibinfo {title} {Self-consistent equations
  including exchange and correlation effects},\ }\href
  {https://doi.org/10.1103/PhysRev.140.A1133} {\bibfield  {journal} {\bibinfo
  {journal} {Phys. Rev.}\ }\textbf {\bibinfo {volume} {140}},\ \bibinfo {pages}
  {1133} (\bibinfo {year} {1965})}\BibitemShut {NoStop}%
\bibitem [{\citenamefont {Kohn}(1999)}]{kohn1999}%
  \BibitemOpen
  \bibfield  {author} {\bibinfo {author} {\bibfnamefont {W.}~\bibnamefont
  {Kohn}},\ }\bibfield  {title} {\bibinfo {title} {Nobel lecture: Electronic
  structure of matter\char22{}wave functions and density functionals},\ }\href
  {https://doi.org/10.1103/RevModPhys.71.1253} {\bibfield  {journal} {\bibinfo
  {journal} {Rev. Mod. Phys.}\ }\textbf {\bibinfo {volume} {71}},\ \bibinfo
  {pages} {1253} (\bibinfo {year} {1999})}\BibitemShut {NoStop}%
\bibitem [{\citenamefont {Tokatly}(2013)}]{tokatly2013}%
  \BibitemOpen
  \bibfield  {author} {\bibinfo {author} {\bibfnamefont {I.~V.}\ \bibnamefont
  {Tokatly}},\ }\bibfield  {title} {\bibinfo {title} {Time-dependent density
  functional theory for many-electron systems interacting with cavity
  photons},\ }\href {https://doi.org/10.1103/PhysRevLett.110.233001} {\bibfield
   {journal} {\bibinfo  {journal} {Phys. Rev. Lett.}\ }\textbf {\bibinfo
  {volume} {110}},\ \bibinfo {pages} {233001} (\bibinfo {year}
  {2013})}\BibitemShut {NoStop}%
\bibitem [{\citenamefont {Ruggenthaler}\ \emph {et~al.}(2014)\citenamefont
  {Ruggenthaler}, \citenamefont {Flick}, \citenamefont {Pellegrini},
  \citenamefont {Appel}, \citenamefont {Tokatly},\ and\ \citenamefont
  {Rubio}}]{ruggenthaler2014}%
  \BibitemOpen
  \bibfield  {author} {\bibinfo {author} {\bibfnamefont {M.}~\bibnamefont
  {Ruggenthaler}}, \bibinfo {author} {\bibfnamefont {J.}~\bibnamefont {Flick}},
  \bibinfo {author} {\bibfnamefont {C.}~\bibnamefont {Pellegrini}}, \bibinfo
  {author} {\bibfnamefont {H.}~\bibnamefont {Appel}}, \bibinfo {author}
  {\bibfnamefont {I.~V.}\ \bibnamefont {Tokatly}},\ and\ \bibinfo {author}
  {\bibfnamefont {A.}~\bibnamefont {Rubio}},\ }\bibfield  {title} {\bibinfo
  {title} {Quantum-electrodynamical density-functional theory: Bridging quantum
  optics and electronic-structure theory},\ }\href
  {https://doi.org/10.1103/PhysRevA.90.012508} {\bibfield  {journal} {\bibinfo
  {journal} {Phys. Rev. A}\ }\textbf {\bibinfo {volume} {90}},\ \bibinfo
  {pages} {012508} (\bibinfo {year} {2014})}\BibitemShut {NoStop}%
\bibitem [{\citenamefont {Rivera}\ \emph {et~al.}(2019)\citenamefont {Rivera},
  \citenamefont {Flick},\ and\ \citenamefont {Narang}}]{rivera2018}%
  \BibitemOpen
  \bibfield  {author} {\bibinfo {author} {\bibfnamefont {N.}~\bibnamefont
  {Rivera}}, \bibinfo {author} {\bibfnamefont {J.}~\bibnamefont {Flick}},\ and\
  \bibinfo {author} {\bibfnamefont {P.}~\bibnamefont {Narang}},\ }\bibfield
  {title} {\bibinfo {title} {Variational theory of nonrelativistic quantum
  electrodynamics},\ }\href {https://doi.org/10.1103/PhysRevLett.122.193603}
  {\bibfield  {journal} {\bibinfo  {journal} {Phys. Rev. Lett.}\ }\textbf
  {\bibinfo {volume} {122}},\ \bibinfo {pages} {193603} (\bibinfo {year}
  {2019})}\BibitemShut {NoStop}%
\bibitem [{\citenamefont {Haugland}\ \emph {et~al.}(2020)\citenamefont
  {Haugland}, \citenamefont {Ronca}, \citenamefont {Kj\o{}nstad}, \citenamefont
  {Rubio},\ and\ \citenamefont {Koch}}]{Haugland2020}%
  \BibitemOpen
  \bibfield  {author} {\bibinfo {author} {\bibfnamefont {T.~S.}\ \bibnamefont
  {Haugland}}, \bibinfo {author} {\bibfnamefont {E.}~\bibnamefont {Ronca}},
  \bibinfo {author} {\bibfnamefont {E.~F.}\ \bibnamefont {Kj\o{}nstad}},
  \bibinfo {author} {\bibfnamefont {A.}~\bibnamefont {Rubio}},\ and\ \bibinfo
  {author} {\bibfnamefont {H.}~\bibnamefont {Koch}},\ }\bibfield  {title}
  {\bibinfo {title} {Coupled cluster theory for molecular polaritons: Changing
  ground and excited states},\ }\href
  {https://doi.org/10.1103/PhysRevX.10.041043} {\bibfield  {journal} {\bibinfo
  {journal} {Phys. Rev. X}\ }\textbf {\bibinfo {volume} {10}},\ \bibinfo
  {pages} {041043} (\bibinfo {year} {2020})}\BibitemShut {NoStop}%
\bibitem [{\citenamefont {Vendrell}(2018)}]{vendrell2018}%
  \BibitemOpen
  \bibfield  {author} {\bibinfo {author} {\bibfnamefont {O.}~\bibnamefont
  {Vendrell}},\ }\bibfield  {title} {\bibinfo {title} {Coherent dynamics in
  cavity femtochemistry: Application of the multi-configuration time-dependent
  hartree method},\ }\href
  {https://doi.org/https://doi.org/10.1016/j.chemphys.2018.02.008} {\bibfield
  {journal} {\bibinfo  {journal} {Chem. Phys.}\ }\textbf {\bibinfo {volume}
  {509}},\ \bibinfo {pages} {55} (\bibinfo {year} {2018})}\BibitemShut
  {NoStop}%
\bibitem [{\citenamefont {Hoffmann}\ \emph {et~al.}(2018)\citenamefont
  {Hoffmann}, \citenamefont {Appel}, \citenamefont {Rubio},\ and\ \citenamefont
  {Maitra}}]{hoffmann2018}%
  \BibitemOpen
  \bibfield  {author} {\bibinfo {author} {\bibfnamefont {N.~M.}\ \bibnamefont
  {Hoffmann}}, \bibinfo {author} {\bibfnamefont {H.}~\bibnamefont {Appel}},
  \bibinfo {author} {\bibfnamefont {A.}~\bibnamefont {Rubio}},\ and\ \bibinfo
  {author} {\bibfnamefont {N.~T.}\ \bibnamefont {Maitra}},\ }\bibfield  {title}
  {\bibinfo {title} {Light-matter interactions via the exact factorization
  approach},\ }\href {https://doi.org/10.1140/epjb/e2018-90177-6} {\bibfield
  {journal} {\bibinfo  {journal} {Eur. Phys. J. B}\ }\textbf {\bibinfo {volume}
  {91}},\ \bibinfo {pages} {180} (\bibinfo {year} {2018})}\BibitemShut
  {NoStop}%
\bibitem [{\citenamefont {Abedi}\ \emph {et~al.}(2018)\citenamefont {Abedi},
  \citenamefont {Khosravi},\ and\ \citenamefont {Tokatly}}]{abedi2018}%
  \BibitemOpen
  \bibfield  {author} {\bibinfo {author} {\bibfnamefont {A.}~\bibnamefont
  {Abedi}}, \bibinfo {author} {\bibfnamefont {E.}~\bibnamefont {Khosravi}},\
  and\ \bibinfo {author} {\bibfnamefont {I.~V.}\ \bibnamefont {Tokatly}},\
  }\bibfield  {title} {\bibinfo {title} {Shedding light on correlated
  electron{\textendash}photon states using the exact factorization},\ }\href
  {https://doi.org/10.1140/epjb/e2018-90243-1} {\bibfield  {journal} {\bibinfo
  {journal} {Eur. Phys. J. B}\ }\textbf {\bibinfo {volume} {91}},\ \bibinfo
  {pages} {194} (\bibinfo {year} {2018})}\BibitemShut {NoStop}%
\bibitem [{\citenamefont {Mordovina}\ \emph {et~al.}(2020)\citenamefont
  {Mordovina}, \citenamefont {Bungey}, \citenamefont {Appel}, \citenamefont
  {Knowles}, \citenamefont {Rubio},\ and\ \citenamefont
  {Manby}}]{Mordovina2020}%
  \BibitemOpen
  \bibfield  {author} {\bibinfo {author} {\bibfnamefont {U.}~\bibnamefont
  {Mordovina}}, \bibinfo {author} {\bibfnamefont {C.}~\bibnamefont {Bungey}},
  \bibinfo {author} {\bibfnamefont {H.}~\bibnamefont {Appel}}, \bibinfo
  {author} {\bibfnamefont {P.~J.}\ \bibnamefont {Knowles}}, \bibinfo {author}
  {\bibfnamefont {A.}~\bibnamefont {Rubio}},\ and\ \bibinfo {author}
  {\bibfnamefont {F.~R.}\ \bibnamefont {Manby}},\ }\bibfield  {title} {\bibinfo
  {title} {Polaritonic coupled-cluster theory},\ }\href
  {https://doi.org/10.1103/PhysRevResearch.2.023262} {\bibfield  {journal}
  {\bibinfo  {journal} {Phys. Rev. Research}\ }\textbf {\bibinfo {volume}
  {2}},\ \bibinfo {pages} {023262} (\bibinfo {year} {2020})}\BibitemShut
  {NoStop}%
\bibitem [{\citenamefont {Haugland}\ \emph {et~al.}(2021)\citenamefont
  {Haugland}, \citenamefont {Schäfer}, \citenamefont {Ronca}, \citenamefont
  {Rubio},\ and\ \citenamefont {Koch}}]{Haugland2021}%
  \BibitemOpen
  \bibfield  {author} {\bibinfo {author} {\bibfnamefont {T.~S.}\ \bibnamefont
  {Haugland}}, \bibinfo {author} {\bibfnamefont {C.}~\bibnamefont {Schäfer}},
  \bibinfo {author} {\bibfnamefont {E.}~\bibnamefont {Ronca}}, \bibinfo
  {author} {\bibfnamefont {A.}~\bibnamefont {Rubio}},\ and\ \bibinfo {author}
  {\bibfnamefont {H.}~\bibnamefont {Koch}},\ }\bibfield  {title} {\bibinfo
  {title} {Intermolecular interactions in optical cavities: An ab initio qed
  study},\ }\href {https://doi.org/10.1063/5.0039256} {\bibfield  {journal}
  {\bibinfo  {journal} {The Journal of Chemical Physics}\ }\textbf {\bibinfo
  {volume} {154}},\ \bibinfo {pages} {094113} (\bibinfo {year}
  {2021})}\BibitemShut {NoStop}%
\bibitem [{\citenamefont {DePrince}(2021)}]{DePrince2021}%
  \BibitemOpen
  \bibfield  {author} {\bibinfo {author} {\bibfnamefont {A.~E.}\ \bibnamefont
  {DePrince}},\ }\bibfield  {title} {\bibinfo {title} {Cavity-modulated
  ionization potentials and electron affinities from quantum electrodynamics
  coupled-cluster theory},\ }\href {https://doi.org/10.1063/5.0038748}
  {\bibfield  {journal} {\bibinfo  {journal} {The Journal of Chemical Physics}\
  }\textbf {\bibinfo {volume} {154}},\ \bibinfo {pages} {094112} (\bibinfo
  {year} {2021})}\BibitemShut {NoStop}%
\bibitem [{\citenamefont {Burke}(2012)}]{burke2012}%
  \BibitemOpen
  \bibfield  {author} {\bibinfo {author} {\bibfnamefont {K.}~\bibnamefont
  {Burke}},\ }\bibfield  {title} {\bibinfo {title} {Perspective on density
  functional theory},\ }\href {https://doi.org/10.1063/1.4704546} {\bibfield
  {journal} {\bibinfo  {journal} {The Journal of Chemical Physics}\ }\textbf
  {\bibinfo {volume} {136}},\ \bibinfo {pages} {150901} (\bibinfo {year}
  {2012})}\BibitemShut {NoStop}%
\bibitem [{\citenamefont {Lehtola}\ \emph {et~al.}(2018)\citenamefont
  {Lehtola}, \citenamefont {Steigemann}, \citenamefont {Oliveira},\ and\
  \citenamefont {Marques}}]{Lehtola_2018}%
  \BibitemOpen
  \bibfield  {author} {\bibinfo {author} {\bibfnamefont {S.}~\bibnamefont
  {Lehtola}}, \bibinfo {author} {\bibfnamefont {C.}~\bibnamefont {Steigemann}},
  \bibinfo {author} {\bibfnamefont {M.~J.}\ \bibnamefont {Oliveira}},\ and\
  \bibinfo {author} {\bibfnamefont {M.~A.}\ \bibnamefont {Marques}},\
  }\bibfield  {title} {\bibinfo {title} {Recent developments in libxc
  {\textemdash} a comprehensive library of functionals for density functional
  theory},\ }\href {https://doi.org/10.1016/j.softx.2017.11.002} {\bibfield
  {journal} {\bibinfo  {journal} {{SoftwareX}}\ }\textbf {\bibinfo {volume}
  {7}},\ \bibinfo {pages} {1} (\bibinfo {year} {2018})}\BibitemShut {NoStop}%
\bibitem [{\citenamefont {Hohenberg}\ and\ \citenamefont
  {Kohn}(1964)}]{hohenberg1964}%
  \BibitemOpen
  \bibfield  {author} {\bibinfo {author} {\bibfnamefont {P.}~\bibnamefont
  {Hohenberg}}\ and\ \bibinfo {author} {\bibfnamefont {W.}~\bibnamefont
  {Kohn}},\ }\bibfield  {title} {\bibinfo {title} {Inhomogeneous electron
  gas},\ }\href {https://doi.org/10.1103/PhysRev.136.B864} {\bibfield
  {journal} {\bibinfo  {journal} {Phys. Rev.}\ }\textbf {\bibinfo {volume}
  {136}},\ \bibinfo {pages} {864} (\bibinfo {year} {1964})}\BibitemShut
  {NoStop}%
\bibitem [{\citenamefont {Perdew}\ and\ \citenamefont
  {Zunger}(1981)}]{perdew1981}%
  \BibitemOpen
  \bibfield  {author} {\bibinfo {author} {\bibfnamefont {J.~P.}\ \bibnamefont
  {Perdew}}\ and\ \bibinfo {author} {\bibfnamefont {A.}~\bibnamefont
  {Zunger}},\ }\bibfield  {title} {\bibinfo {title} {Self-interaction
  correction to density-functional approximations for many-electron systems},\
  }\href {https://doi.org/10.1103/PhysRevB.23.5048} {\bibfield  {journal}
  {\bibinfo  {journal} {Phys. Rev. B}\ }\textbf {\bibinfo {volume} {23}},\
  \bibinfo {pages} {5048} (\bibinfo {year} {1981})}\BibitemShut {NoStop}%
\bibitem [{\citenamefont {Becke}(1988)}]{Becke1988}%
  \BibitemOpen
  \bibfield  {author} {\bibinfo {author} {\bibfnamefont {A.~D.}\ \bibnamefont
  {Becke}},\ }\bibfield  {title} {\bibinfo {title} {Density-functional
  exchange-energy approximation with correct asymptotic behavior},\ }\href
  {https://doi.org/10.1103/PhysRevA.38.3098} {\bibfield  {journal} {\bibinfo
  {journal} {Phys. Rev. A}\ }\textbf {\bibinfo {volume} {38}},\ \bibinfo
  {pages} {3098} (\bibinfo {year} {1988})}\BibitemShut {NoStop}%
\bibitem [{\citenamefont {Perdew}\ \emph {et~al.}(1996)\citenamefont {Perdew},
  \citenamefont {Burke},\ and\ \citenamefont {Ernzerhof}}]{perdew1996}%
  \BibitemOpen
  \bibfield  {author} {\bibinfo {author} {\bibfnamefont {J.~P.}\ \bibnamefont
  {Perdew}}, \bibinfo {author} {\bibfnamefont {K.}~\bibnamefont {Burke}},\ and\
  \bibinfo {author} {\bibfnamefont {M.}~\bibnamefont {Ernzerhof}},\ }\bibfield
  {title} {\bibinfo {title} {Generalized gradient approximation made simple},\
  }\href {https://doi.org/10.1103/PhysRevLett.77.3865} {\bibfield  {journal}
  {\bibinfo  {journal} {Phys. Rev. Lett.}\ }\textbf {\bibinfo {volume} {77}},\
  \bibinfo {pages} {3865} (\bibinfo {year} {1996})}\BibitemShut {NoStop}%
\bibitem [{\citenamefont {Heyd}\ \emph {et~al.}(2003)\citenamefont {Heyd},
  \citenamefont {Scuseria},\ and\ \citenamefont {Ernzerhof}}]{heyd2003}%
  \BibitemOpen
  \bibfield  {author} {\bibinfo {author} {\bibfnamefont {J.}~\bibnamefont
  {Heyd}}, \bibinfo {author} {\bibfnamefont {G.~E.}\ \bibnamefont {Scuseria}},\
  and\ \bibinfo {author} {\bibfnamefont {M.}~\bibnamefont {Ernzerhof}},\
  }\bibfield  {title} {\bibinfo {title} {Hybrid functionals based on a screened
  coulomb potential},\ }\href {https://doi.org/10.1063/1.1564060} {\bibfield
  {journal} {\bibinfo  {journal} {The Journal of Chemical Physics}\ }\textbf
  {\bibinfo {volume} {118}},\ \bibinfo {pages} {8207} (\bibinfo {year}
  {2003})}\BibitemShut {NoStop}%
\bibitem [{\citenamefont {Krukau}\ \emph {et~al.}(2006)\citenamefont {Krukau},
  \citenamefont {Vydrov}, \citenamefont {Izmaylov},\ and\ \citenamefont
  {Scuseria}}]{Krukau2006}%
  \BibitemOpen
  \bibfield  {author} {\bibinfo {author} {\bibfnamefont {A.~V.}\ \bibnamefont
  {Krukau}}, \bibinfo {author} {\bibfnamefont {O.~A.}\ \bibnamefont {Vydrov}},
  \bibinfo {author} {\bibfnamefont {A.~F.}\ \bibnamefont {Izmaylov}},\ and\
  \bibinfo {author} {\bibfnamefont {G.~E.}\ \bibnamefont {Scuseria}},\
  }\bibfield  {title} {\bibinfo {title} {Influence of the exchange screening
  parameter on the performance of screened hybrid functionals},\ }\href
  {https://doi.org/10.1063/1.2404663} {\bibfield  {journal} {\bibinfo
  {journal} {The Journal of Chemical Physics}\ }\textbf {\bibinfo {volume}
  {125}},\ \bibinfo {pages} {224106} (\bibinfo {year} {2006})}\BibitemShut
  {NoStop}%
\bibitem [{\citenamefont {K\"ummel}\ and\ \citenamefont
  {Kronik}(2008)}]{kuemmel2008}%
  \BibitemOpen
  \bibfield  {author} {\bibinfo {author} {\bibfnamefont {S.}~\bibnamefont
  {K\"ummel}}\ and\ \bibinfo {author} {\bibfnamefont {L.}~\bibnamefont
  {Kronik}},\ }\bibfield  {title} {\bibinfo {title} {Orbital-dependent density
  functionals: Theory and applications},\ }\href
  {https://doi.org/10.1103/RevModPhys.80.3} {\bibfield  {journal} {\bibinfo
  {journal} {Rev. Mod. Phys.}\ }\textbf {\bibinfo {volume} {80}},\ \bibinfo
  {pages} {3} (\bibinfo {year} {2008})}\BibitemShut {NoStop}%
\bibitem [{\citenamefont {Pellegrini}\ \emph {et~al.}(2015)\citenamefont
  {Pellegrini}, \citenamefont {Flick}, \citenamefont {Tokatly}, \citenamefont
  {Appel},\ and\ \citenamefont {Rubio}}]{pellegrini2015}%
  \BibitemOpen
  \bibfield  {author} {\bibinfo {author} {\bibfnamefont {C.}~\bibnamefont
  {Pellegrini}}, \bibinfo {author} {\bibfnamefont {J.}~\bibnamefont {Flick}},
  \bibinfo {author} {\bibfnamefont {I.~V.}\ \bibnamefont {Tokatly}}, \bibinfo
  {author} {\bibfnamefont {H.}~\bibnamefont {Appel}},\ and\ \bibinfo {author}
  {\bibfnamefont {A.}~\bibnamefont {Rubio}},\ }\bibfield  {title} {\bibinfo
  {title} {Optimized effective potential for quantum electrodynamical
  time-dependent density functional theory},\ }\href
  {https://doi.org/10.1103/PhysRevLett.115.093001} {\bibfield  {journal}
  {\bibinfo  {journal} {Phys. Rev. Lett.}\ }\textbf {\bibinfo {volume} {115}},\
  \bibinfo {pages} {093001} (\bibinfo {year} {2015})}\BibitemShut {NoStop}%
\bibitem [{\citenamefont {Flick}\ \emph {et~al.}(2018)\citenamefont {Flick},
  \citenamefont {Schäfer}, \citenamefont {Ruggenthaler}, \citenamefont
  {Appel},\ and\ \citenamefont {Rubio}}]{flick2018a}%
  \BibitemOpen
  \bibfield  {author} {\bibinfo {author} {\bibfnamefont {J.}~\bibnamefont
  {Flick}}, \bibinfo {author} {\bibfnamefont {C.}~\bibnamefont {Schäfer}},
  \bibinfo {author} {\bibfnamefont {M.}~\bibnamefont {Ruggenthaler}}, \bibinfo
  {author} {\bibfnamefont {H.}~\bibnamefont {Appel}},\ and\ \bibinfo {author}
  {\bibfnamefont {A.}~\bibnamefont {Rubio}},\ }\bibfield  {title} {\bibinfo
  {title} {Ab initio optimized effective potentials for real molecules in
  optical cavities: Photon contributions to the molecular ground state},\
  }\href {https://doi.org/10.1021/acsphotonics.7b01279} {\bibfield  {journal}
  {\bibinfo  {journal} {ACS Photonics}\ }\textbf {\bibinfo {volume} {5}},\
  \bibinfo {pages} {992} (\bibinfo {year} {2018})}\BibitemShut {NoStop}%
\bibitem [{\citenamefont {Vydrov}\ and\ \citenamefont
  {Van~Voorhis}(2010)}]{vydrov2010}%
  \BibitemOpen
  \bibfield  {author} {\bibinfo {author} {\bibfnamefont {O.~A.}\ \bibnamefont
  {Vydrov}}\ and\ \bibinfo {author} {\bibfnamefont {T.}~\bibnamefont
  {Van~Voorhis}},\ }\bibfield  {title} {\bibinfo {title} {Dispersion
  interactions from a local polarizability model},\ }\href
  {https://doi.org/10.1103/PhysRevA.81.062708} {\bibfield  {journal} {\bibinfo
  {journal} {Phys. Rev. A}\ }\textbf {\bibinfo {volume} {81}},\ \bibinfo
  {pages} {062708} (\bibinfo {year} {2010})}\BibitemShut {NoStop}%
\bibitem [{\citenamefont {Ruggenthaler}(2017)}]{ruggenthaler2017groundstate}%
  \BibitemOpen
  \bibfield  {author} {\bibinfo {author} {\bibfnamefont {M.}~\bibnamefont
  {Ruggenthaler}},\ }\href@noop {} {\bibinfo {title} {Ground-state
  quantum-electrodynamical density-functional theory}} (\bibinfo {year}
  {2017}),\ \Eprint {https://arxiv.org/abs/1509.01417} {arXiv:1509.01417
  [quant-ph]} \BibitemShut {NoStop}%
\bibitem [{\citenamefont {Flick}\ \emph {et~al.}(2015)\citenamefont {Flick},
  \citenamefont {Ruggenthaler}, \citenamefont {Appel},\ and\ \citenamefont
  {Rubio}}]{flick2015}%
  \BibitemOpen
  \bibfield  {author} {\bibinfo {author} {\bibfnamefont {J.}~\bibnamefont
  {Flick}}, \bibinfo {author} {\bibfnamefont {M.}~\bibnamefont {Ruggenthaler}},
  \bibinfo {author} {\bibfnamefont {H.}~\bibnamefont {Appel}},\ and\ \bibinfo
  {author} {\bibfnamefont {A.}~\bibnamefont {Rubio}},\ }\bibfield  {title}
  {\bibinfo {title} {Kohn{\textendash}sham approach to quantum electrodynamical
  density-functional theory: Exact time-dependent effective potentials in real
  space},\ }\href {https://doi.org/10.1073/pnas.1518224112} {\bibfield
  {journal} {\bibinfo  {journal} {Proceedings of the National Academy of
  Sciences}\ }\textbf {\bibinfo {volume} {112}},\ \bibinfo {pages} {15285}
  (\bibinfo {year} {2015})}\BibitemShut {NoStop}%
\bibitem [{\citenamefont {Harris}(1984)}]{harris1984}%
  \BibitemOpen
  \bibfield  {author} {\bibinfo {author} {\bibfnamefont {J.}~\bibnamefont
  {Harris}},\ }\bibfield  {title} {\bibinfo {title} {Adiabatic-connection
  approach to kohn-sham theory},\ }\href
  {https://doi.org/10.1103/PhysRevA.29.1648} {\bibfield  {journal} {\bibinfo
  {journal} {Phys. Rev. A}\ }\textbf {\bibinfo {volume} {29}},\ \bibinfo
  {pages} {1648} (\bibinfo {year} {1984})}\BibitemShut {NoStop}%
\bibitem [{\citenamefont {Heßelmann}\ and\ \citenamefont
  {G\"orling}(2011)}]{hesselmann2011}%
  \BibitemOpen
  \bibfield  {author} {\bibinfo {author} {\bibfnamefont {A.}~\bibnamefont
  {Heßelmann}}\ and\ \bibinfo {author} {\bibfnamefont {A.}~\bibnamefont
  {G\"orling}},\ }\bibfield  {title} {\bibinfo {title} {Random-phase
  approximation correlation methods for molecules and solids},\ }\href
  {https://doi.org/10.1080/00268976.2011.614282} {\bibfield  {journal}
  {\bibinfo  {journal} {Molecular Physics}\ }\textbf {\bibinfo {volume}
  {109}},\ \bibinfo {pages} {2473} (\bibinfo {year} {2011})}\BibitemShut
  {NoStop}%
\bibitem [{\citenamefont {Flick}\ \emph {et~al.}(2019)\citenamefont {Flick},
  \citenamefont {Welakuh}, \citenamefont {Ruggenthaler}, \citenamefont
  {Appel},\ and\ \citenamefont {Rubio}}]{flick2019}%
  \BibitemOpen
  \bibfield  {author} {\bibinfo {author} {\bibfnamefont {J.}~\bibnamefont
  {Flick}}, \bibinfo {author} {\bibfnamefont {D.~M.}\ \bibnamefont {Welakuh}},
  \bibinfo {author} {\bibfnamefont {M.}~\bibnamefont {Ruggenthaler}}, \bibinfo
  {author} {\bibfnamefont {H.}~\bibnamefont {Appel}},\ and\ \bibinfo {author}
  {\bibfnamefont {A.}~\bibnamefont {Rubio}},\ }\bibfield  {title} {\bibinfo
  {title} {Light–matter response in nonrelativistic quantum
  electrodynamics},\ }\href {https://doi.org/10.1021/acsphotonics.9b00768}
  {\bibfield  {journal} {\bibinfo  {journal} {ACS Photonics}\ }\textbf
  {\bibinfo {volume} {6}},\ \bibinfo {pages} {2757} (\bibinfo {year}
  {2019})}\BibitemShut {NoStop}%
\bibitem [{\citenamefont {Hermann}\ \emph {et~al.}(2017)\citenamefont
  {Hermann}, \citenamefont {DiStasio},\ and\ \citenamefont
  {Tkatchenko}}]{Hermann2017}%
  \BibitemOpen
  \bibfield  {author} {\bibinfo {author} {\bibfnamefont {J.}~\bibnamefont
  {Hermann}}, \bibinfo {author} {\bibfnamefont {R.~A.}\ \bibnamefont
  {DiStasio}},\ and\ \bibinfo {author} {\bibfnamefont {A.}~\bibnamefont
  {Tkatchenko}},\ }\bibfield  {title} {\bibinfo {title} {First-principles
  models for van der waals interactions in molecules and materials: Concepts,
  theory, and applications},\ }\href
  {https://doi.org/10.1021/acs.chemrev.6b00446} {\bibfield  {journal} {\bibinfo
   {journal} {Chemical Reviews}\ }\textbf {\bibinfo {volume} {117}},\ \bibinfo
  {pages} {4714} (\bibinfo {year} {2017})}\BibitemShut {NoStop}%
\bibitem [{\citenamefont {Wang}\ \emph {et~al.}(2021)\citenamefont {Wang},
  \citenamefont {Neuman}, \citenamefont {Flick},\ and\ \citenamefont
  {Narang}}]{wang2021}%
  \BibitemOpen
  \bibfield  {author} {\bibinfo {author} {\bibfnamefont {D.~S.}\ \bibnamefont
  {Wang}}, \bibinfo {author} {\bibfnamefont {T.}~\bibnamefont {Neuman}},
  \bibinfo {author} {\bibfnamefont {J.}~\bibnamefont {Flick}},\ and\ \bibinfo
  {author} {\bibfnamefont {P.}~\bibnamefont {Narang}},\ }\bibfield  {title}
  {\bibinfo {title} {Light–matter interaction of a molecule in a dissipative
  cavity from first principles},\ }\href {https://doi.org/10.1063/5.0036283}
  {\bibfield  {journal} {\bibinfo  {journal} {The Journal of Chemical Physics}\
  }\textbf {\bibinfo {volume} {154}},\ \bibinfo {pages} {104109} (\bibinfo
  {year} {2021})}\BibitemShut {NoStop}%
\bibitem [{\citenamefont {H\"ummer}\ \emph {et~al.}(2013)\citenamefont
  {H\"ummer}, \citenamefont {Garc\'{\i}a-Vidal}, \citenamefont
  {Mart\'{\i}n-Moreno},\ and\ \citenamefont {Zueco}}]{hummer2013}%
  \BibitemOpen
  \bibfield  {author} {\bibinfo {author} {\bibfnamefont {T.}~\bibnamefont
  {H\"ummer}}, \bibinfo {author} {\bibfnamefont {F.~J.}\ \bibnamefont
  {Garc\'{\i}a-Vidal}}, \bibinfo {author} {\bibfnamefont {L.}~\bibnamefont
  {Mart\'{\i}n-Moreno}},\ and\ \bibinfo {author} {\bibfnamefont
  {D.}~\bibnamefont {Zueco}},\ }\bibfield  {title} {\bibinfo {title} {Weak and
  strong coupling regimes in plasmonic qed},\ }\href
  {https://doi.org/10.1103/PhysRevB.87.115419} {\bibfield  {journal} {\bibinfo
  {journal} {Phys. Rev. B}\ }\textbf {\bibinfo {volume} {87}},\ \bibinfo
  {pages} {115419} (\bibinfo {year} {2013})}\BibitemShut {NoStop}%
\bibitem [{\citenamefont {Scheel}\ and\ \citenamefont
  {Buhmann}(2008)}]{scheel2009macroscopic}%
  \BibitemOpen
  \bibfield  {author} {\bibinfo {author} {\bibfnamefont {S.}~\bibnamefont
  {Scheel}}\ and\ \bibinfo {author} {\bibfnamefont {S.~Y.}\ \bibnamefont
  {Buhmann}},\ }\bibfield  {title} {\bibinfo {title} {Macroscopic qed-concepts
  and applications},\ }\href@noop {} {\bibfield  {journal} {\bibinfo  {journal}
  {acta physica slovaca}\ ,\ \bibinfo {pages} {675}} (\bibinfo {year}
  {2008})}\BibitemShut {NoStop}%
\bibitem [{\citenamefont {Keller}(2012)}]{keller2012quantum}%
  \BibitemOpen
  \bibfield  {author} {\bibinfo {author} {\bibfnamefont {O.}~\bibnamefont
  {Keller}},\ }\href {https://books.google.com/books?id=v2ck\_\_wFOBEC} {\emph
  {\bibinfo {title} {Quantum Theory of Near-Field Electrodynamics}}},\
  Nano-Optics and Nanophotonics\ (\bibinfo  {publisher} {Springer Berlin
  Heidelberg},\ \bibinfo {year} {2012})\BibitemShut {NoStop}%
\bibitem [{\citenamefont {Venkataram}\ \emph {et~al.}(2017)\citenamefont
  {Venkataram}, \citenamefont {Hermann}, \citenamefont {Tkatchenko},\ and\
  \citenamefont {Rodriguez}}]{venkataram2017}%
  \BibitemOpen
  \bibfield  {author} {\bibinfo {author} {\bibfnamefont {P.~S.}\ \bibnamefont
  {Venkataram}}, \bibinfo {author} {\bibfnamefont {J.}~\bibnamefont {Hermann}},
  \bibinfo {author} {\bibfnamefont {A.}~\bibnamefont {Tkatchenko}},\ and\
  \bibinfo {author} {\bibfnamefont {A.~W.}\ \bibnamefont {Rodriguez}},\
  }\bibfield  {title} {\bibinfo {title} {Unifying microscopic and continuum
  treatments of van der waals and casimir interactions},\ }\href
  {https://doi.org/10.1103/PhysRevLett.118.266802} {\bibfield  {journal}
  {\bibinfo  {journal} {Phys. Rev. Lett.}\ }\textbf {\bibinfo {volume} {118}},\
  \bibinfo {pages} {266802} (\bibinfo {year} {2017})}\BibitemShut {NoStop}%
\bibitem [{\citenamefont {Wylie}\ and\ \citenamefont {Sipe}(1985)}]{wylie1985}%
  \BibitemOpen
  \bibfield  {author} {\bibinfo {author} {\bibfnamefont {J.~M.}\ \bibnamefont
  {Wylie}}\ and\ \bibinfo {author} {\bibfnamefont {J.~E.}\ \bibnamefont
  {Sipe}},\ }\bibfield  {title} {\bibinfo {title} {Quantum electrodynamics near
  an interface. ii},\ }\href {https://doi.org/10.1103/PhysRevA.32.2030}
  {\bibfield  {journal} {\bibinfo  {journal} {Phys. Rev. A}\ }\textbf {\bibinfo
  {volume} {32}},\ \bibinfo {pages} {2030} (\bibinfo {year}
  {1985})}\BibitemShut {NoStop}%
\bibitem [{\citenamefont {Marques}\ \emph {et~al.}(2003)\citenamefont
  {Marques}, \citenamefont {Castro}, \citenamefont {Bertsch},\ and\
  \citenamefont {Rubio}}]{octopus1}%
  \BibitemOpen
  \bibfield  {author} {\bibinfo {author} {\bibfnamefont {M.~A.~L.}\
  \bibnamefont {Marques}}, \bibinfo {author} {\bibfnamefont {A.}~\bibnamefont
  {Castro}}, \bibinfo {author} {\bibfnamefont {G.~F.}\ \bibnamefont
  {Bertsch}},\ and\ \bibinfo {author} {\bibfnamefont {A.}~\bibnamefont
  {Rubio}},\ }\bibfield  {title} {\bibinfo {title} {{Octopus: a
  First-Principles Tool for Excited Electron-Ion Dynamics}},\ }\href
  {https://www.sciencedirect.com/science/article/pii/S0010465502006860}
  {\bibfield  {journal} {\bibinfo  {journal} {Comput. Phys. Commun.}\ }\textbf
  {\bibinfo {volume} {151}},\ \bibinfo {pages} {60} (\bibinfo {year}
  {2003})}\BibitemShut {NoStop}%
\bibitem [{\citenamefont {Andrade}\ \emph {et~al.}(2015)\citenamefont
  {Andrade}, \citenamefont {A.}, \citenamefont {De~Giovannini}, \citenamefont
  {Larsen}, \citenamefont {Oliveira}, \citenamefont {Alberdi-Rodriguez},
  \citenamefont {Varas}, \citenamefont {Theophilou}, \citenamefont {Helbig},
  \citenamefont {Verstraete}, \citenamefont {Stella}, \citenamefont {Nogueira},
  \citenamefont {Aspuru-Guzik}, \citenamefont {Castro}, \citenamefont
  {Marques},\ and\ \citenamefont {Rubio}}]{octopus2}%
  \BibitemOpen
  \bibfield  {author} {\bibinfo {author} {\bibfnamefont {X.}~\bibnamefont
  {Andrade}}, \bibinfo {author} {\bibfnamefont {S.~D.}\ \bibnamefont {A.}},
  \bibinfo {author} {\bibfnamefont {U.}~\bibnamefont {De~Giovannini}}, \bibinfo
  {author} {\bibfnamefont {A.~H.}\ \bibnamefont {Larsen}}, \bibinfo {author}
  {\bibfnamefont {M.~J.~T.}\ \bibnamefont {Oliveira}}, \bibinfo {author}
  {\bibfnamefont {J.}~\bibnamefont {Alberdi-Rodriguez}}, \bibinfo {author}
  {\bibfnamefont {A.}~\bibnamefont {Varas}}, \bibinfo {author} {\bibfnamefont
  {I.}~\bibnamefont {Theophilou}}, \bibinfo {author} {\bibfnamefont
  {N.}~\bibnamefont {Helbig}}, \bibinfo {author} {\bibfnamefont {M.~J.}\
  \bibnamefont {Verstraete}}, \bibinfo {author} {\bibfnamefont
  {L.}~\bibnamefont {Stella}}, \bibinfo {author} {\bibfnamefont
  {F.}~\bibnamefont {Nogueira}}, \bibinfo {author} {\bibfnamefont
  {A.}~\bibnamefont {Aspuru-Guzik}}, \bibinfo {author} {\bibfnamefont
  {A.}~\bibnamefont {Castro}}, \bibinfo {author} {\bibfnamefont {M.~A.~L.}\
  \bibnamefont {Marques}},\ and\ \bibinfo {author} {\bibfnamefont
  {A.}~\bibnamefont {Rubio}},\ }\bibfield  {title} {\bibinfo {title}
  {{Real-Space Grids and the Octopus Code as Tools for the Development of New
  Simulation Approaches for Electronic Systems}},\ }\href
  {https://pubs.rsc.org/en/content/articlelanding/2015/cp/c5cp00351b#!divAbstract}
  {\bibfield  {journal} {\bibinfo  {journal} {Phys. Chem. Chem. Phys.}\
  }\textbf {\bibinfo {volume} {17}},\ \bibinfo {pages} {31371} (\bibinfo {year}
  {2015})}\BibitemShut {NoStop}%
\bibitem [{\citenamefont {Tancogne-Dejean}\ \emph {et~al.}(2020)\citenamefont
  {Tancogne-Dejean}, \citenamefont {Oliveira}, \citenamefont {Andrade},
  \citenamefont {Appel}, \citenamefont {Borca}, \citenamefont {Le~Breton},
  \citenamefont {Buchholz}, \citenamefont {Castro}, \citenamefont {Corni},
  \citenamefont {Correa}, \citenamefont {De~Giovannini}, \citenamefont
  {Delgado}, \citenamefont {Eich}, \citenamefont {Flick}, \citenamefont {Gil},
  \citenamefont {Gomez}, \citenamefont {Helbig}, \citenamefont {Hübener},
  \citenamefont {Jestädt} \emph {et~al.}}]{octopus3}%
  \BibitemOpen
  \bibfield  {author} {\bibinfo {author} {\bibfnamefont {N.}~\bibnamefont
  {Tancogne-Dejean}}, \bibinfo {author} {\bibfnamefont {M.~J.~T.}\ \bibnamefont
  {Oliveira}}, \bibinfo {author} {\bibfnamefont {X.}~\bibnamefont {Andrade}},
  \bibinfo {author} {\bibfnamefont {H.}~\bibnamefont {Appel}}, \bibinfo
  {author} {\bibfnamefont {C.~H.}\ \bibnamefont {Borca}}, \bibinfo {author}
  {\bibfnamefont {G.}~\bibnamefont {Le~Breton}}, \bibinfo {author}
  {\bibfnamefont {F.}~\bibnamefont {Buchholz}}, \bibinfo {author}
  {\bibfnamefont {A.}~\bibnamefont {Castro}}, \bibinfo {author} {\bibfnamefont
  {S.}~\bibnamefont {Corni}}, \bibinfo {author} {\bibfnamefont {A.~A.}\
  \bibnamefont {Correa}}, \bibinfo {author} {\bibfnamefont {U.}~\bibnamefont
  {De~Giovannini}}, \bibinfo {author} {\bibfnamefont {A.}~\bibnamefont
  {Delgado}}, \bibinfo {author} {\bibfnamefont {F.~G.}\ \bibnamefont {Eich}},
  \bibinfo {author} {\bibfnamefont {J.}~\bibnamefont {Flick}}, \bibinfo
  {author} {\bibfnamefont {G.}~\bibnamefont {Gil}}, \bibinfo {author}
  {\bibfnamefont {A.}~\bibnamefont {Gomez}}, \bibinfo {author} {\bibfnamefont
  {N.}~\bibnamefont {Helbig}}, \bibinfo {author} {\bibfnamefont
  {H.}~\bibnamefont {Hübener}}, \bibinfo {author} {\bibfnamefont
  {R.}~\bibnamefont {Jestädt}}, \emph {et~al.},\ }\bibfield  {title} {\bibinfo
  {title} {Octopus, a computational framework for exploring light-driven
  phenomena and quantum dynamics in extended and finite systems},\ }\href
  {https://doi.org/10.1063/1.5142502} {\bibfield  {journal} {\bibinfo
  {journal} {J. Chem. Phys.}\ }\textbf {\bibinfo {volume} {152}},\ \bibinfo
  {pages} {124119} (\bibinfo {year} {2020})}\BibitemShut {NoStop}%
\bibitem [{\citenamefont {Troullier}\ and\ \citenamefont
  {Martins}(1991)}]{troullier1993}%
  \BibitemOpen
  \bibfield  {author} {\bibinfo {author} {\bibfnamefont {N.}~\bibnamefont
  {Troullier}}\ and\ \bibinfo {author} {\bibfnamefont {J.~L.}\ \bibnamefont
  {Martins}},\ }\bibfield  {title} {\bibinfo {title} {Efficient
  pseudopotentials for plane-wave calculations},\ }\href
  {https://doi.org/10.1103/PhysRevB.43.1993} {\bibfield  {journal} {\bibinfo
  {journal} {Phys. Rev. B}\ }\textbf {\bibinfo {volume} {43}},\ \bibinfo
  {pages} {1993} (\bibinfo {year} {1991})}\BibitemShut {NoStop}%
\end{thebibliography}%

\appendix
\section{Appendix}

\subsection{(A) Adiabatic Connection}

We can introduce the adiabatic connection between the physical interacting system and the non-interacting Kohn-Sham system with similar arguments as in the electronic system, e.g. Refs.~\cite{harris1984,hesselmann2011}. We can do so by considering a Hamiltonian of the following form
\begin{align*}
\hat{H}_\gamma = \hat{T}_e + \hat{H}_p + \gamma \hat{H}_\text{int} + \hat{H}^{\gamma}_\text{ext} 
\end{align*}
where the dimensionless parameter $\gamma \in [0,1]$. This Hamiltonian $\hat{H}_\gamma$ recovers in the limit of $\gamma=1$ the physical Hamiltonian of Eq.~\ref{eqn:Hfull}. In the opposite limit with $\gamma=0$, we find a non-interacting Hamiltonian. Now let $\hat{H}^{\gamma}_\text{ext}$ consist of an arbitrary one-electron potential and photonic current. In addition, we assume the state $\ket{\Psi_0}$ is an eigenstate of the non-interacting Hamiltonian $\hat{H}_0$. This eigenstate defines the corresponding internal variables $n_0 (\textbf{r}) = \bra{\Psi_0} \hat{n}(\textbf{r}) \ket{\Psi_0}$ and $q_{\alpha,0} = \bra{\Psi_0} \hat{q}_\alpha \ket{\Psi_0}$. In general, $\ket{\Psi_0}$ will be a product of a single Slater determinant with quantum harmonic oscillator eigenstates. The parameter $\gamma$ is now increased from $0$ to $d\gamma$ while keeping the external variables constant. The perturbation $d\gamma \hat{H}_\text{int}$ will change $\ket{\Psi_0}$ to $\ket{\Psi_0'}$, an eigenstate of $\hat{H}'_0 = \hat{H}_0 + d\gamma \hat{H}_\text{int}$, that corresponds to new internal variables $(n'_0,q'_{\alpha,0})$ . It is now assumed that an additional perturbation in the form of an external field, $d\gamma \hat{H}^{0'}_\text{ext}$ can be found such that the state $\ket{\Psi'_0}$ becomes $\ket{\Psi_{d\lambda}}$, which is an eigenstate of 
\begin{align*}
\hat{H}_{d\gamma} = \hat{H}_0 + d\gamma \left(\hat{H}_\text{int} + \hat{H}^{0'}_\text{ext} \right)
\end{align*}
and leads to the internal variables $n_0$ and  $q_{\alpha,0}$. The conditions that such an external Hamiltonian exists are identical to regular DFT~\cite{harris1984,hesselmann2011}.

If we now combine these perturbations the original eigenstate $\ket{\Psi_0}$ goes into eigenstate $\ket{\Psi_{d\lambda}}$ of Hamiltonian $\hat{H}_{d\lambda}$ at constant internal variables. By continuing this procedure in infinitesimal steps going from $\gamma=0$ to $\gamma=1$, we have defined a smooth adiabatic path with constant internal variables $n$ and $q$, but where the eigenstate $\ket{\Psi_0}$ becomes the eigenstate $\ket{\Psi_1}$ of the interacting Hamiltonian $\hat{H}_1=\hat H$ with the physical values for electron-electron and electron-photon interaction.

\subsection{(B) Connection to OEP}
\label{app:oep}

In this section, we show that Eq.~\ref{eq:exc2} and ~\ref{eq:exc1} lead to the photon OEP energy expression introduced in Ref.~\cite{pellegrini2015} (Eq. 12). As a first step, we can use Eq.~\ref{eq:exc2} and make use of the following relation~\cite{wylie1985}
\begin{align*}
    \int_0^\infty \frac{d\epsilon}{\left(\epsilon^2 + a^2 \right)\left(\epsilon^2 + b^2 \right)} = \frac{\pi}{2 |a||b| \left( |a| + |b|\right)}
\end{align*}
leading to 
\begin{align*}
    E_c^{(1,1)} =  -\frac{1}{2}\sum_\alpha\sum_{ia} \frac{\omega_\alpha|d_{ia}|^2}{\epsilon_a-\epsilon_i + \omega_\alpha}
\end{align*}
with $d_{ia}=\bra{\varphi_i}\boldsymbol\lambda_\alpha\cdot \textbf r \ket{\varphi_a}$, where $i$ and $a$ are occupied and unoccupied Kohn-Sham orbitals, respectively. This expression for $E_{c}^{(1,1)}$ is exactly the first part of the OEP energy in Ref.~\cite{pellegrini2015} (Eq. 12). The second part can be obtained by using Eq.~\ref{eq:exc1} and the relation 
\begin{align*}
    \int_0^\infty dx \, \frac{1}{b^2+x^2} = \frac{\pi}{2|b|}
\end{align*}
leading to
\begin{align*}
    E_x = \frac{1}{2}\sum_\alpha\sum_{ia} {\omega_\alpha|d_{ia}|^2}.
\end{align*}
Finally, $E_x^{(p)}=E_x+E_c^{(1,1)}$ leads to the full expression of the photon OEP energy in Ref.~\cite{pellegrini2015} (Eq. 12).

\subsection{(C) Computational details}
\label{app:comp}
We have implemented the developed scheme into the pseudopotential, real-space time-dependent density functional theory (TDDFT) code Octopus~\cite{octopus1, octopus2,octopus3}. For the calculations shown in this paper, we setup a real-space grid using spheres of 6\AA~around each atom with a grid spacing of 0.15\AA.
While we explicitly describe all valence electrons in our simulations on a real-space grid, we describe the core-electrons with pseudopotentials and use the LDA Troullier-Martins pseudopotentials~\cite{troullier1993}. All relaxations haven been performed with the LDA exchange-correlation functional~\cite{perdew1981} until forces on the individual atoms are smaller than $5\times10^{-3}$ eV/\AA. For the calculations in Fig.~\ref{fig:01}, for the KLI/OEP calculations, we describe electrons and photons self-consistent on the level of KLI/OEP. For OEP-ss and GA calclulations, we treat the electrons self-consistent on the level of OEP and add the OEP/GA energy non-self-consistently in the last iteration. For the calculations shown in Fig.~\ref{fig:02}, we treat the electronic structure on the level of LDA~\cite{perdew1981} and add the photon GA energy in the last iteration non-self-consistently. 

\end{document}